\documentclass[aps,amsmath,amssymb,preprintnumbers,a4paper,prd,twocolumn]{revtex4-1}
 \pdfoutput=1

 \usepackage{lipsum}

\usepackage{slashed}
\usepackage{subfigure}
\usepackage{graphicx}
\usepackage{amsfonts}
\usepackage{color}
\usepackage{cancel}
\def\eq#1{{eq.~(\ref{#1})}}

\def\fig#1{{fig.~(\ref{#1})}}

\def\fig#1{{fig.~(\ref{#1})}}

\def\hbar{\hspace{0pt}\raisebox{1pt}{$-$} \hspace{-7pt} h}

\def\5{\overline 5}

\definecolor{JJ}{RGB}{0,144,255}

\newcommand{\be}{\begin{equation}}
\newcommand{\ee}{\end{equation}}
\newcommand{\bea}{\begin{eqnarray}}
\newcommand{\eea}{\end{eqnarray}}

\newcommand{\ba}{\begin{eqnarray}}
\newcommand{\ea}{\end{eqnarray}}

\newcommand{\GeV}{\mbox{ ${\mathrm{GeV}}$}}
\newcommand{\TeV}{\mbox{ ${\mathrm{TeV}}$}}

\begin{document}
\title{Symmetries and composite dynamics for the 750 GeV diphoton excess}

\author{Diogo Buarque Franzosi}
 \affiliation{
 II. Physikalisches Institut, Universit\"at G\"ottingen, Friedrich-Hund-Platz 1, 37077 G\"ottingen, Germany
}
\author{Mads T. Frandsen}
 \affiliation{CP$^{3}$-Origins and the Danish IAS, University of Southern Denmark, Campusvej 55, DK-5230 Odense M, Denmark}

\begin{abstract} 

The ATLAS and CMS experiments at LHC observe small excesses of diphoton events with invariant mass around 750 GeV. Here we study the possibility of nearly parity degenerate and vector-scalar degenerate spectra as well as composite dynamics in 2 scenarios for explaining the excess: Production of a pseudo-scalar via gluon or photon fusion or via decay of a parent particle together with soft additional final states.  
We discuss possible underlying realizations of the scenarios motivated by dynamical models of electroweak symmetry breaking (without new coloured states) and fermion masses.

\preprint{CP3-Origins-2015-023 DNRF90, DIAS-2015-23}

\end{abstract}

\maketitle

 The ATLAS  and CMS \cite{ATLAS,CMS:2015dxe} experiments observe local excesses in diphoton final states with invariant masses around 750 GeV. If these excesses are interpreted as a narrow resonance the local significances correspond to 3.6 and 2.6 $\sigma$ respectively leading to modest global significances of 2$\sigma$ in ATLAS and 1.2$\sigma$ in CMS. The significance is slightly higher in the ATLAS experiment if interpreted as a resonance with a width of about 50 GeV while the CMS significance is decreased slightly. The results suggest a reconstructed resonance mass $P$ of around $750$~GeV with a production cross-section $\sigma_P \sim 5-10 ~{\rm fb} $.
Although the significances of the results are rather modest and have been challenged in e.g~\cite{Davis:2016hlw} it is well motivated to entertain the idea that these results may be due to new 
physics.

General phenomenological analyses indicate that the new state is of composite nature~\cite{Franceschini:2015kwy,*Ellis:2015oso,*Gupta:2015zzs,*Appelquist:2016viq}, even though other interpretations in terms of simple extensions of the Standard Model (SM)~\cite{Falkowski:2015swt,*Ghosh:2015apa,*Han:2015qqj, *Hernandez:2015ywg,*Moretti:2015pbj,*Badziak:2015zez}, Supersymmetric models~\cite{Petersson:2015mkr,*Ding:2015rxx,*Feng:2015wil,*Chakraborty:2015gyj},  Goldstones, Radions and Dilatons~\cite{Bellazzini:2015nxw,*Cox:2015ckc,*Ahmed:2015uqt,*Megias:2015ory,*Bardhan:2015hcr}, axions~\cite{Molinaro:2015cwg} as well as a wide variety of models or interpretations ~\cite{Nomura:2016seu,*Faraggi:2016xnm,
*Djouadi:2016eyy,
*Dorsner:2016ypw,
*Laperashvili:2016cah,
*Fodor:2016wal,
*Ding:2016ldt,
*Yu:2016lof,
*Hati:2016thk,
*Arun:2016ela,
*Cao:2016udb,
*Ko:2016wce,
*Xia:2016jec,
*Sonmez:2016xov,
*Borah:2016uoi,
*Sahin:2016lda,
*Bhattacharya:2016lyg,
*D'Eramo:2016mgv,
*Berlin:2016hqw,
*Zhang:2016pip,
*Ito:2016zkz,
*Dutta:2016jqn,
*Modak:2016ung,
*Csaki:2016raa,
*Chao:2016mtn,
*Danielsson:2016nyy,
*Ko:2016lai,
*Nomura:2016fzs,
*Palti:2016kew,
*Potter:2016psi,
*Ghorbani:2016jdq,
*Han:2016bvl,
*Aad:2015owa,
*CMS:2015gla,
*Jung:2015etr,
*Dasgupta:2015pbr,
*Kaneta:2015qpf,
*Jiang:2015oms,
*Hernandez:2015hrt,
*Low:2015qho,
*Dong:2015dxw,
*Kanemura:2015bli,
*Kang:2015roj,
*Chiang:2015tqz,
*Huang:2015svl,
*Hamada:2015skp,
*Pich:2015lkh,
*Bizot:2015qqo,
*Anchordoqui:2015jxc,
*Bi:2015lcf,
*Chao:2015nac,
*Kim:2015xyn,
*Cai:2015hzc,
*Cao:2015apa,
*Tang:2015eko,
*Dev:2015vjd,
*Gao:2015igz,
*Cao:2015scs,
*Wang:2015omi,
*Son:2015vfl,
*Li:2015jwd,
*Salvio:2015jgu,
*DelDebbio:2015byq,
*Chway:2015lzg,
*Park:2015ysf,
*Han:2015yjk,
*Hall:2015xds,
*Casas:2015blx,
*Zhang:2015uuo,
*Liu:2015yec,
*Cheung:2015cug,
*Das:2015enc,
*Davoudiasl:2015cuo,
*Allanach:2015ixl,
*Cvetic:2015vit,
*Patel:2015ulo,
*Gu:2015lxj,
*Cao:2015xjz,
*Chabab:2015nel,
*Huang:2015rkj,
*Pelaggi:2015knk,
*Dey:2015bur,
*Dev:2015isx,
*Kulkarni:2015gzu,
*Cline:2015msi,
*Berthier:2015vbb,
*Kim:2015ksf,
*Bi:2015uqd,
*Cao:2015twy,
*Wang:2015kuj,
*Antipin:2015kgh,
*Hatanaka:2015qjo,
*Chakraborty:2015jvs,
*Barducci:2015gtd,
*Chang:2015sdy,
*Luo:2015yio,
*Han:2015dlp,
*Dhuria:2015ufo,
*Chang:2015bzc,
*Han:2015cty,
*Arun:2015ubr,
*Ringwald:2015dsf,
*Chao:2015nsm,
*Bernon:2015abk,
*Alves:2015jgx,
*Bai:2015nbs,
*Bian:2015kjt,
*Curtin:2015jcv,
*Chao:2015ttq,
*Demidov:2015zqn,
*No:2015bsn,
*Becirevic:2015fmu,
*Agrawal:2015dbf,
*Kobakhidze:2015ldh,
*Dutta:2015wqh,
*Low:2015qep,
*McDermott:2015sck,
*DiChiara:2015vdm,
*Buttazzo:2015txu,
*Biswas:2015zgk,
*Chiang:2015amq,
*Harigaya:2015ezk,
*Marzola:2015xbh,
*deBlas:2015hlv,
*Heckman:2015kqk,
*Cao:2015pto,
*Buckley2016,
*Chiang2016a,
*Okada2016,
*Ibanez2015,
*Kanemura2015,
*Cho2015,
*Diaz2015,
*Gopalakrishna2015,
*Sun2014}, have been contemplated. 
Implications for cosmology and dark matter have been discussed \emph{e.g.} in~\cite{Mambrini:2015wyu,*Backovic:2015fnp,*Chao:2016aer}.

\bigskip 
In this study we present 2 scenarios for the diphoton excess motivated by composite dynamics for EWSB and spectral symmetries.  In particular we examine if the signal can be achieved without invoking new coloured states. Specific models of strong dynamics in which the new states carry color have been studied in~\cite{Harigaya2016,*Nakai:2015ptz,*Matsuzaki:2015che}.

In {\it scenario 1} a pseudo scalar $P$ with a large diphoton partial width is produced in gluon fusion via a top Yukawa of order $5\times 10^{-2} \lesssim y_{Pt}\lesssim 0.7$. The upper limit is chosen so that $\Gamma_{P\to tt}/m_P\lesssim 0.05$, and the total width of $P$ is near the best fit width from the ATLAS data. If $P$ is a composite of new EW charged strongly interacting fermions, involved in EWSB but with a decay constant below $v_{\rm EW}$, a sufficiently large diphoton decay rate is possible in minimal models.  

In {\it scenario 2} the pseudo-scalar $P$ is produced via decays of a parent resonance $R$ (see \emph{e.g.}~\cite{Kim:2015ron,*Huang:2015evq,*An:2015cgp}) with a small mass splitting $\delta m_{RP}=m_R - m_P$. Accordingly $P$ does not need to couple significantly to SM fermions and the branching to diphotons can be $O(1)$. The small mass splitting ensures that the additional final state particles can be soft. Finally $P$ can be off-shell when $\delta m_{RP}<0$,  mimicking a finite width in the diphoton decays. If $P$ and $R$ are composite states of EW charged strongly interacting fermions, involved in EWSB,  the $O(1)$ diphoton branching ratio can be achieved and the small mass splitting realized by symmetries. 
For example parity doubling of the spectrum in the large-$N$ limit of the underlying gauge group or via a vector-scalar symmetry such as in the heavy quark limit of QCD.
This scenario, with negligible couplings of $P$ to SM fermions  allows the opposite extreme of  scenario 1.

We outline specific underlying composite dynamics, motivated by dynamical EWSB and fermion mass generation, able to provide the features needed to explain the excess in both scenarios. 
The experimental signature distinguishing {\it scenario 1} and {\it scenario 2}, independent of the specific realizations we discuss are the additional soft final states which may be difficult to detect. 

\bigskip

To motivate the first scenario we note that for an $s$-channel pseudo-scalar resonance $P$ coupled to gluons and photons the excess and the best-fit width in the ATLAS data are reproduced for $\Gamma_{P}/m_P\sim 0.05$ and $\Gamma_{\gamma\gamma}\Gamma_{gg}/m_P^2\sim 5 \times 10^{-8}$ \cite{Franceschini:2015kwy}. For a top Yukawa of $0.7$ we have $\Gamma_{gg}/m_P\sim  10^{-4}$ and $\Gamma_{P\to tt}/m_P\sim 0.05$. For a simple 2 Dirac flavor model of dynamical EWSB \cite{Dietrich:2005jn} with an additional lepton doublet and a low compositeness scale $f_P \lesssim 0.4 v_{\rm EW}$ we find  $\Gamma_{\gamma\gamma}/m_P\simeq 2 \times 10^{-4}$ not too far from the required values. Moreover the ratio $m_P/f_P \sim 10$ for $m_P=750$ GeV and $f_P=75$ GeV is similar to the ratio $m_{\eta'}/f_\pi$ in QCD and not unnatural.

\bigskip
As an alternative we consider that the (pseudo-) scalar resonance is produced in the decay $R \to X P ( \to \gamma\gamma)$ of a new 'parent' resonance $R$ either of spin-0 ($S$) or of spin-1 $(V)$. The LHC diphoton analysis are inclusive so there could be interesting additional activity $X$ along with the diphoton system, but in this study we require the additional final state(s) $X$ to be soft. 

For the spin-1 parent $V$ we take a neutral isosinglet vector resonance with a mass splitting, $\delta m_{VP}=m_V - m_P$, relative to $P$ and coupled predominantly to $b$-quarks. The signal process is $\sigma (pp(\bar{b}b) \to V \to P X \to \gamma\gamma \, X)$ with $X=\{\gamma_{\rm soft}, Z^*\}$ a soft additional photon or off-shell $Z$ boson. Having dominant couplings to $b$-quarks are preferred by the constraints on diphoton resonances from the 8 TeV run of LHC which favor a high increase in production cross-section from 8 to 13 TeV.  

For the case of a spin-0 parent the scalar resonance $S$ may be endowed with sizable top-quark couplings e.g. via mixing with the SM Higgs while fermion couplings may be negligible for the pseudo scalar $P$ --- allowing a large diphoton branching ratio for $P$. 

Even if this might imply a narrow width for $P$ it can appear as a resonance with finite width if produced slightly off-shell \cite{An:2015cgp} when $\delta m_{SP}=m_S - m_P$ is small and negative.

We motivate the three scenarios by outlining composite models of dynamical electroweak symmetry breaking and fermion masses that may realize them. 
The pseudo scalar $P$ may be interpreted as a resonance analogous to the $\pi^0, \eta$ and $\eta'$ states in QCD with sizable diphoton branching ratios. 
The $V,S$ parent resonances can be thought of as analogues of e.g. the $\omega$ and $\sigma$ mesons in QCD.  

The small mass splittings $\delta m_{VP}, \delta m_{SP}$ can e.g. be realized in composite dynamics via parity doubling of the spectrum in the large-$N$ limit of the underlying gauge group or via a vector-scalar symmetry such as in the heavy quark limit of QCD.  For certain fermion representations we indeed expect  $\delta m_{SP}$ to be small and negative in the large-N limit, allowing off-shell production  \cite{Sannino:2003xe}.

Moreover, isosinglet  '$\omega$-like' vector resonances $V$ will, for some underlying realizations, not mass mix with the SM weak spin-1 bosons and only be coupled to SM fermions via higher dimensional operators \cite{Chivukula:1989rn}. See appendix of \cite{Foadi:2007ue} for an explicit example.  The higher dimensional operators may naturally induce dominant couplings to the 3rd generation SM quarks, for example if these operators are related to a dynamical SM fermion mass generation mechanism like the so-called 'Extended Technicolor' construction \cite{Eichten:1979ah,Dimopoulos:1979es}. 
 
\bigskip 
{\it \bf Simplified Lagrangians and production cross-sections:}

We are interested in the signal processes 
\bea
&\sigma& (pp \to P\to 2\gamma) \ , \quad  \nonumber
\\
&\sigma& (pp \to R \to P X \to \gamma\gamma X) \ , \quad 
\label{eq:scenarios}
\eea
with $R$ either a spin-1 resonance $V$ or a scalar resonance $S$ and $X$ denoting additional (soft) final states.

The interaction Lagrangian linking the parent resonances to $P$ and to the standard model fermions can be split into the relevant pieces:
\bea
{\cal L}_{V}  &= &\sum_{f}  \bar{f} \slashed{{V}} \left(g^V_{f }-g^A_{f}\, \gamma_5\right) f  + \frac{1}{\Lambda_V}V_{\mu\nu} \widetilde{F}^{\mu\nu} P\ ,
 \\
 {\cal L}_{P}  &=& i y_{Pf} \, P \bar{f} \gamma^5 f + \frac{1}{\Lambda_P} P F_{\mu\nu} \widetilde{F}^{\mu\nu} 
\\
 {\cal L}_{S } &=&  y_{Sf} \,  S \bar{f}  f + \Lambda_S S P X  
\label{eq:lag}
\eea
 where $f$ is a SM fermion, $V_{\mu\nu}= \partial_\mu V_\nu - \partial_\nu V_\mu$ and $\widetilde{F}^{\mu\nu} =\frac{1}{2}\epsilon^{\mu\nu\rho\sigma}F_{\rho\sigma}$ with $F^{\mu\nu}$ being the field strength of the photon.

The partial widths from the above interactions are, ignoring light fermion masses and the mass of $X$, 
\bea
\Gamma_{P,S\to \bar{f}f} &\simeq&   \frac{N_c(f) }{8\pi} y_{P,S f}^2\,  m_P 
\\
\Gamma_{V\to \bar{f}f} &\simeq& \frac{N_c(f) }{12\pi}
[(g^V_{f })^2+(g^A_{f})^2] m_V
\\
\Gamma_{P\to \gamma\gamma} &\simeq& \frac{m_P^3}{4 \pi \Lambda_P^2}
\\
\Gamma_{V\to P\gamma} &=& \frac{1}{24\pi \Lambda_V^2}\left(\frac{m_V^2-m_P^2}{m_V}\right)^3
\label{Eq:partialsV}
\\
\Gamma_{S\to P X} &\simeq&  \frac{\Lambda_S^2}{16\pi} \frac{m_S^2-m_P^2}{m_S^3}
\label{Eq:partials}
\eea
where $N_c$ is the number of colors of the fermion.

\begin{figure}
\includegraphics[width=0.5\textwidth]{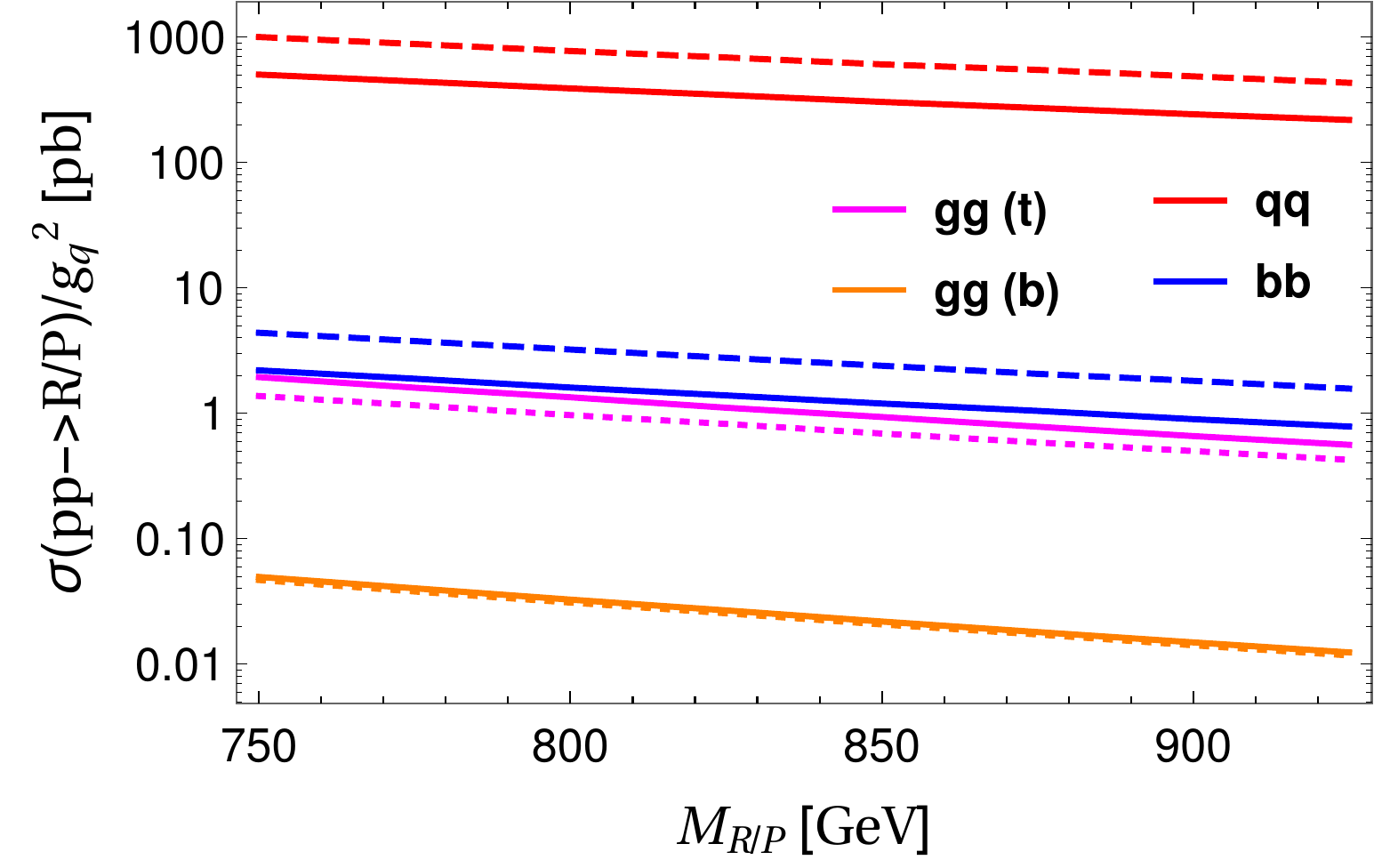}
\caption{Production cross sections of the pseudo-scalar $P$, scalar $S$ and vector $V$ in units of the square of the fermion couplings. 
The label gg refers to loop induced production of (pseudo-) scalars, through a top-quark (t) or a bottom-quark (b) loop. 
The solid lines refer to production of the pseudo-scalar $P$ (identical to $S$ in the quark initiated processes), dashed lines refer to $V$ production and dotted lines refer to $S$ production through gluon fusion.}
\label{fig:xs}
\end{figure}

To fit the peak cross-section of the diphoton excess we assume $m_P\simeq 750$ GeV and require $\delta m_{RP} < 200\GeV$, allowing for some additional final state activity given the inclusive searches. 
The production cross sections in units of the coupling squared are given in Fig.~\ref{fig:xs}, in particular the production of $P$ and $S$ via top-induced gluon fusion (solid purple) and the $b$-quark initiated production of $V$ (blue dashed).
The cross sections are derived at leading order with the factorization and renormalization scales $\mu_{F,R}$ equal to the mass of the produced particle $\mu_F=\mu_R=m_{R,P}$ and the NN23LO1 parton distributions sets \cite{Ball:2012cx}. The loop induced processes have been computed with the model \cite{loopmodel}. Next-to-leading order $K$-factors have been computed for $m_{R,P}=750\GeV$ using the MadGraph\_aMC@NLO program. In the 5 flavour scheme for $b$-quark initiated processes, we obtained $K(bb)=0.99$, see ~\cite{Harlander:2003ai,Wiesemann:2014ioa}. For the light quark initiated processes we used the SM UFO model and increased the mass of the $Z$-boson, obtaining $K(qq)=1.16$. We used the same $K$-factor for the scalar case for simplicity. For the gluon fusion topology we used our own effective gluon-Higgs model implementation and obtained a factor of 2.3. However, we adopted $K(gg)=2.7$ which reproduces the central value of the cross section reported in \cite{Dittmaier:2011ti} at NNLO+NNLL accuracy.
 
  \bigskip 
 {\it \bf Scenarios for the excess}
 
 Based on the above simplified Lagrangian we now detail the two scenarios considered here.
 
In {\it scenario 1} the excess arises from production of the pseudo-scalar $P$ via top-quark loop induced gluon fusion production or photon fusion with the same mechanism of its subsequent decay into photons.
If $P$ couples dominantly to top-quarks in the SM and there are no new colored states, the cross section is well approximated by 
\bea
\sigma_{\gamma\gamma}&\simeq &(\sigma_{gg\to P, 0} \, y_{Pt}^2 +
\sigma_{\gamma\gamma\to P , 0}
\frac{\Gamma_{P\to \gamma\gamma}}{\Gamma_{\gamma\gamma, 0 }} )
\\
&\times&
\frac{\Gamma_{P\to \gamma\gamma}}{\Gamma_{P\to \gamma\gamma}+\Gamma_{P\to VV}+\Gamma_{P\to tt}}  \ , 
\eea
where $\sigma_{gg\to P, 0}=1.9$ pb is a reference cross-section with $y_{Pt}=1$,  $\sigma_{\gamma\gamma\to P , 0}\simeq 3$ fb is a reference cross section for photon-fusion with  $\Gamma_{\gamma\gamma , 0}=0.34$ GeV being the diphoton width for $\Lambda_P=10$ TeV. $P$ will in general decay to weak bosons, $VV=WW,\,ZZ,\,\gamma Z$ with $\Gamma_{P\to VV}= r\Gamma_{P\to\gamma\gamma}$ and $r\gtrsim 1.64$~\cite{Fichet:2015vvy} unless isospin breaking couplings are introduced. We will use $r=1.64$ as a benchmark value. The size of $r$ is constrained by other searches.
We have neglected $\Gamma_{P\to gg}$. 
The  photon fusion contribution to the cross section  is important for small SM fermion couplings. We adopted a naive approximation, using MadGraph with the photon distribution function of the proton. The error on this piece can be large and we refer the reader to recent \cite{Fichet:2015vvy,*Csaki:2015vek,*Sahin:2016lda,*Fichet:2016pvq,*Harland-Lang:2016apc} and on-going \cite{Molinaro} studies for more details.

\begin{figure}
\includegraphics[width=0.5\textwidth]{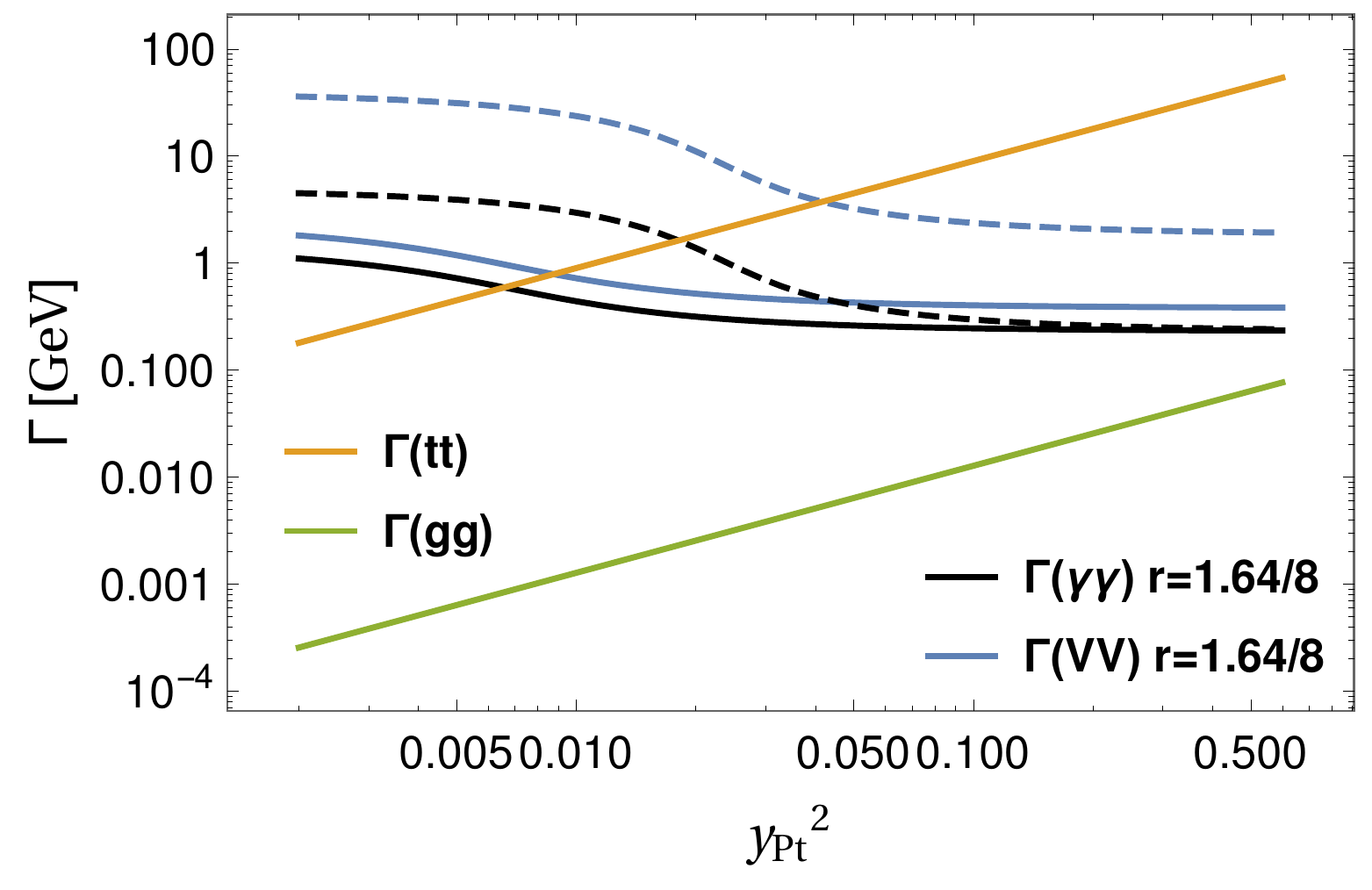}
\caption{
Required partial width $\Gamma_{P\to \gamma\gamma}$ to explain the LHC diphoton excess, $\sigma_{\gamma\gamma}\sim 5$ fb  from gluon fusion production of $P$ via a top-loop and photon fusion, as a function of the Yuakawa coupling $y_{Pt}^2$. The solid  and dashed lines refer to $r=1.64$ and $r=8$ respectively.}
\label{fig:S1ytXGamma}
\end{figure}

We show the required value of $\Gamma_{P\to \gamma\gamma}$ as a function of $y_{Pt}^2$ in \fig{fig:S1ytXGamma} in order to reproduce $\sigma_{\gamma\gamma}\geq 5$ fb.
$\Gamma_{P\to tt}$ and the top loop induced $\Gamma_{P\to gg}$ are also shown.
To reproduce the signal cross-section $\sigma_{\gamma\gamma}\sim 5-10$ fb and the width $\Gamma_P \sim 0.05 m_P$, marginally preferred by the ATLAS results over a narrow resonance (locally 3.6 vs 3.9 $\sigma$), we need $y_{Pt}^2 \sim 0.5$ leading to $\Gamma_{P\to \gamma\gamma}/m_P \sim 3-6 \times 10^{-4}$. The $t\bar{t}$ production via $P\to \bar{t}t$ constrains $y_{Pt}^2\lesssim 1.5$~\cite{Chatrchyan:2013lca}, even though interference effects between signal and background not taken into account in the analysis should relax this bound. Alternatively, for low $y_{Pt}^2$, the large width can be achieved for $r\sim 8$, also shown in \fig{fig:S1ytXGamma}.

The parameter targets in this scenario, in order to avoid experimental constraints, are 
\begin{itemize}
\item $2.5\times 10^{-3} \gtrsim y_{Pt}^2\lesssim 0.6$ \\ 
($3 \times 10^{-4} \gtrsim \Gamma_{P\to tt}/m_P\lesssim 0.05$)
\item   $\Gamma_{P\to \gamma\gamma}/m_P\gtrsim 3-6 \times 10^{-4}$ \ . 
\end{itemize}

 \bigskip 

 In {\it scenario 2} we first consider the Drell-Yan production of a neutral isosinglet spin-1 resonance $R=V$ with subsequent decay $V\to P\gamma$.
 Accordingly the branching ratio ${\rm Br}[P\to \gamma\gamma]\sim O(1)$ is in principle possible as $P$ need not couple significantly to SM fermions to be produced. We adopt a minimum decay rate to weak bosons with $r=1.64$ as discussed before. Additional final state particles can be soft if the mass splitting $\delta m_{VP}$ is sufficiently small. 
 
To get the minimal required production cross-section $\sigma_{pp\to V} \sim 5$ fb, the minimum value of $g_b^V$ or $g_{u,d}^V$ depend on the mass of the parent resonance $V$ and is 
 $g_b^2\gtrsim 1 (3)\times 10^{-3}$ for $m_V=750 (925)$ GeV and $(g_{u,d}^V)^2 \simeq 0 $
 or 
$(g_{u,d}^V)^2 \gtrsim 5 \times 10^{-6} - 10^{-5} $.  The latter is easily achieved via mixing of $V$ with the hyper charge field.
However the former situation is preferred by the constraints from the 8 TeV LHC run since the $b$-initiated cross-section grows with nearly a factor 5 from the 8 to 13 TeV run. 
 
\begin{figure}
\includegraphics[width=0.5\textwidth]{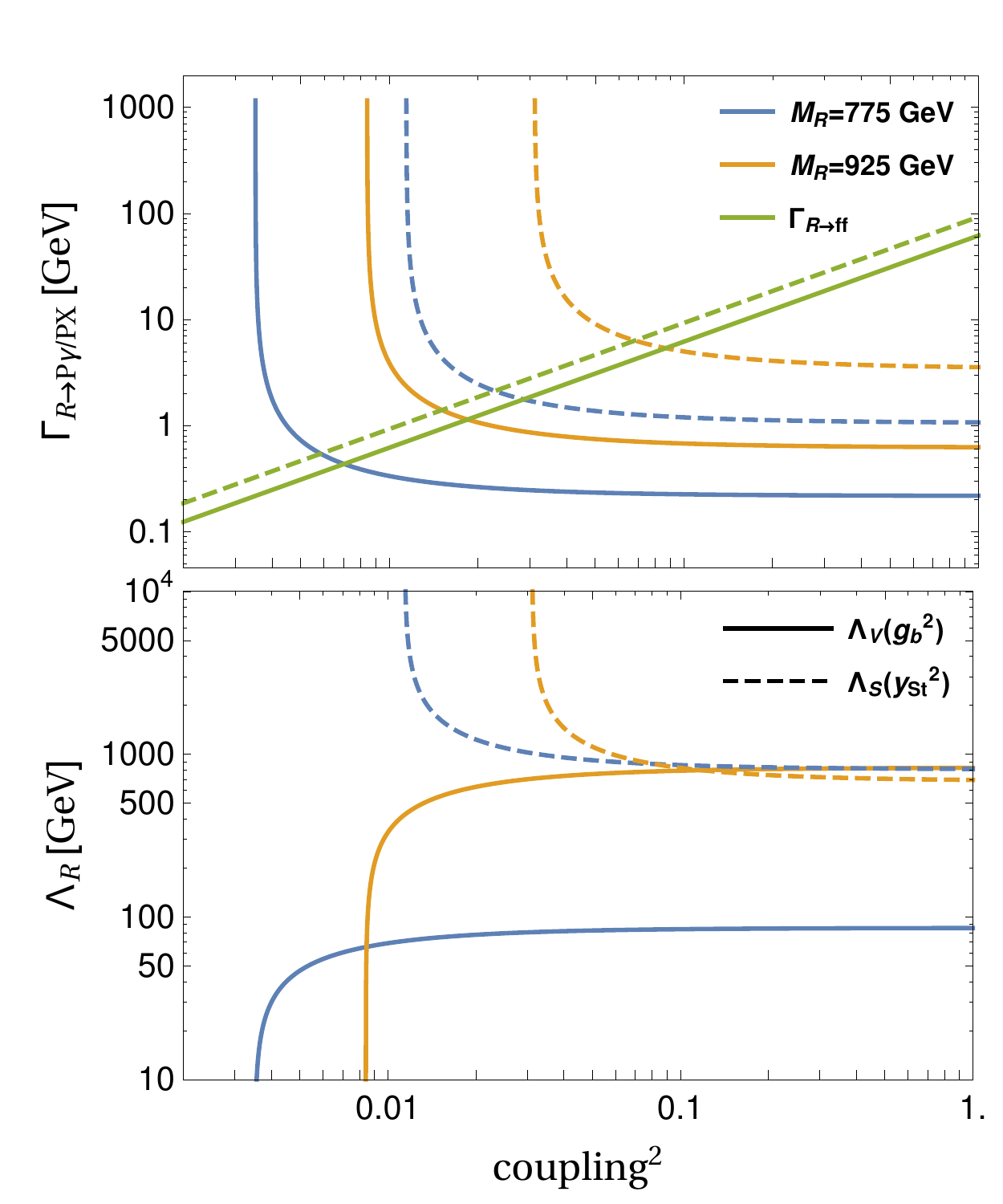}
\caption{\emph{Top:} The partial widths $\Gamma_{S\to PX}$ and $\Gamma_{V\to P\gamma}$ required to explain the di-photon excess as a function of the Yukawa coupling, $y_{St}$ and the vector-fermion coupling $g_b^V$. Also shown the partial width to top(bottom)-quark pair in the S (V) case for $m_R=775\GeV$. \emph{Bottom:} The corresponding scales $\Lambda_S$, $\Lambda_V$ derived from \eq{Eq:partials}.}
\label{fig:LambdaXyt}
\end{figure}

We therefore focus on the $b$-dominated case. Near the minimum possible value of the $b$-coupling  $(g_b^V)^2 \simeq 10^{-3}$ we need 
${\rm Br}[V \to P \gamma_{\rm soft}\to  \gamma\gamma\,  \gamma_{\rm soft}] \simeq 1$. In \fig{fig:LambdaXyt} the required partial width $\Gamma_{V\to P\gamma}$ and the  corresponding values of $\Lambda_V$ are shown:
As $g_b^V$ is increased the needed branching ratio rapidly decrease to a near constant value. 

\begin{figure}
\includegraphics[width=0.5\textwidth]{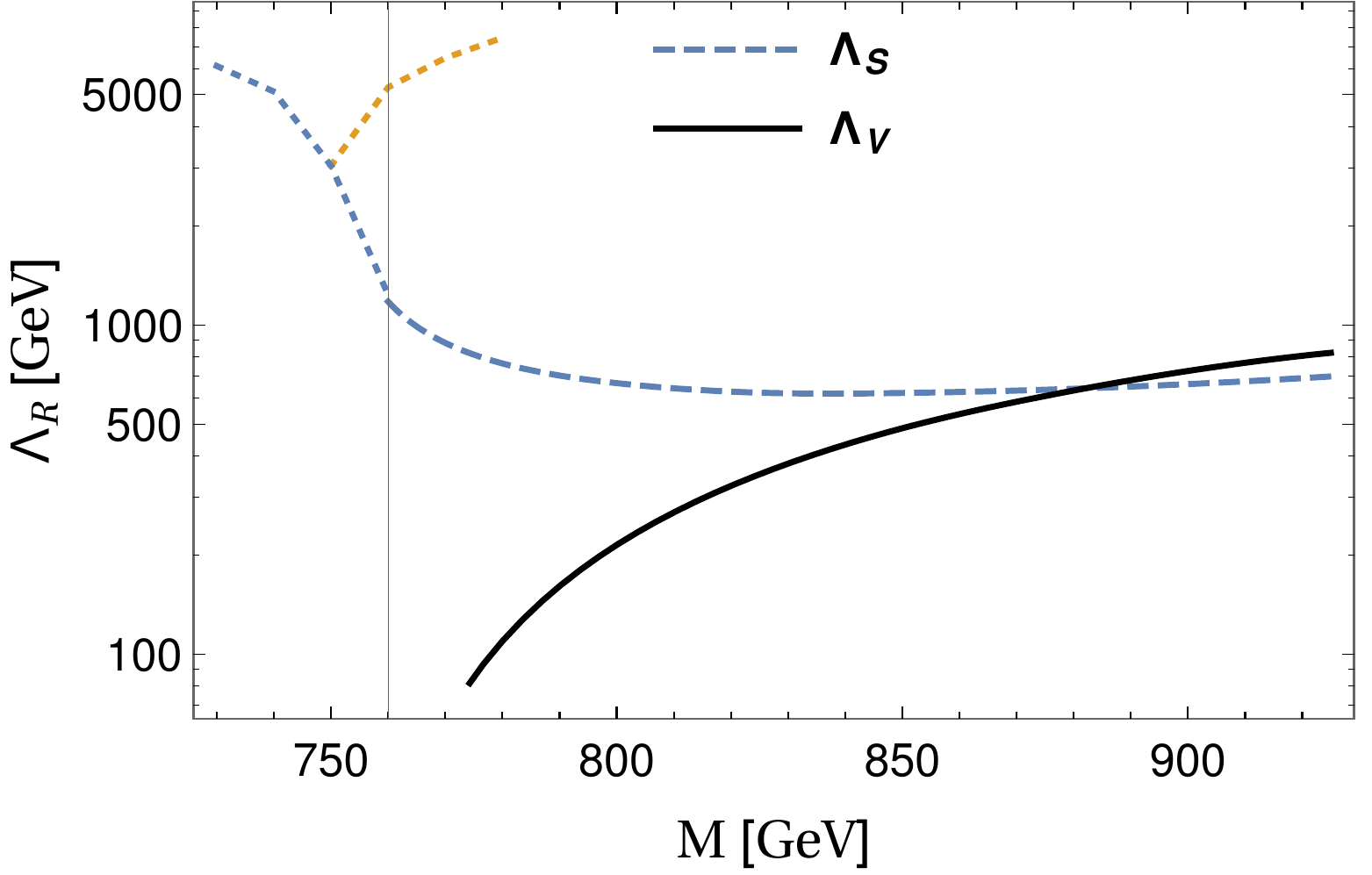}
\caption{The scale $\Lambda_R $ in front of the $R PX$ vertices  required to fit the diphoton excess, with either $R=V$ a spin-1 resonance (black solid), or $R=S$ a scalar resonance (dashed). The scale is  independent of $g_b^V$ for $(g_b^V)^2\gtrsim 3 \times 10^{-2}$ (as seen in the lower panel of fig.~\ref{fig:LambdaXyt} ). The dotted curve correspond to the off-shell region with the orange dotted curve referring to case B, along which $m_P$ increases and $m_S=750\GeV$ is kept fix.}
\label{fig:mvXLV}
\end{figure}

For large values of $g_b^V$ the branching to $b\bar{b}$ is likely to dominate. If that is the case, i.e. ${\rm Br}[V \to bb]> {\rm Br}[V \to P\gamma]$, then since the production coupling grows like $(g_b^V)^2$ 
the required branching to $P\gamma$ becomes $(g_b^V)^2$ independent and only depends on $m_V$ due to the phase space suppression in Eq.~\ref{Eq:partials} and parton distribution suppression. Specifically 
we require  $\Gamma_{R\to P\gamma}\eqsim 0.22-0.63\GeV$ (or equivalently $\Lambda_V\eqsim 85-820 \GeV$) for $m_V=775-925\GeV$.  
In Fig.~\ref{fig:mvXLV} we show the range of relevant scales $\Lambda_R$ as a function of $m_R$ and with $m_P=750$ GeV.
As is evident, the range is very sensitive to the phase space suppression $\delta m_{RP}$.

For large $m_V$, di-jet searches impose the limits $(g_b^V)^2\lesssim 0.6(0.8)$ for $m_V=900(800)\GeV$ \cite{Aad:2014aqa}. 

Finally the above discussion assumed a maximum branching ratio of $P$ to di-photons ${\rm Br}[P \to \gamma \gamma]=1/(1+r)=0.38$  with ($r$=1.64) and negligible couplings of $P$ to SM fermions. 
In specific models that last assumption may not be true, and in one discussed below, indeed $P$ decays also to $b$-quark pairs.

\bigskip 

In scenario 2 the parent resonance may also be a scalar resonance $S$ in a (nearly) parity doubled spectrum of scalar(s) $S$ and pseudo-scalar(s) $P$. 
The scalar $S$ may then e.g. be endowed with top-quark couplings via mixing with the SM Higgs while $P$ can have negligible mixing with the Higgs due to parity and negligible couplings to fermions. Finally the decay $S\to P X$ can be large and dominate $S\to tt$ even for sizable Yukawa couplings. 
The final states $X$ could include soft QCD processes, (parameterized eg. via a mixing of $P$ with pseudo-scalars like the $\eta$ and $\pi$ ), or additional new but light composite states that mix with P as in \cite{Frandsen:2011kt}. 

In order to reproduce the di-photon signal in this scenario the minimal production cross section, $\sigma_{pp\to S}\sim 5$ fb, demands a  Yukawa coupling $y_{Pt}^2 \gtrsim   4 \times 10^{-3}$, for $m_S=775\GeV$, while for $m_S=925\GeV$, $y_{Pt}^2 \gtrsim   0.01 $. The required partial widths and $\Lambda_S$ as a function of $y_{Pt}^2$ are also shown, with dashed lines, in \fig{fig:LambdaXyt}. The decay width to fermions is also shown for comparison.   Different from the vector scenario, the effect of the kinetic suppression is milder and the different resonance mass converge to similar $\Lambda_S$ values for large Yukawa, shown in \fig{fig:mvXLV}.

\begin{figure}
\includegraphics[width=0.5\textwidth]{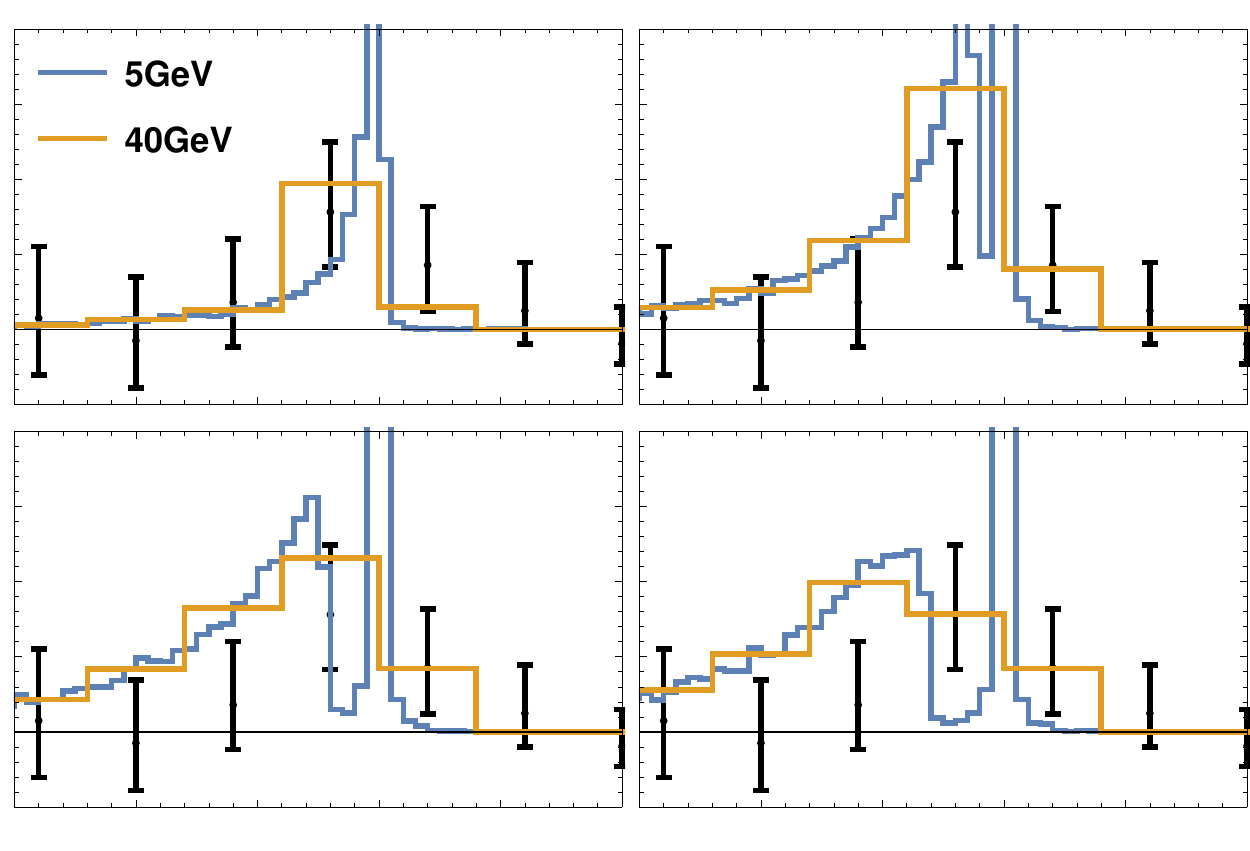}
\caption{Case A: Diphoton distribution from the process $pp\to S \to P X$ with $m_P=m_S=750\GeV$ and $m_X=0,10,20,30\GeV$.}
\label{fig:offshellA}
\end{figure}

\begin{figure}
\includegraphics[width=0.5\textwidth]{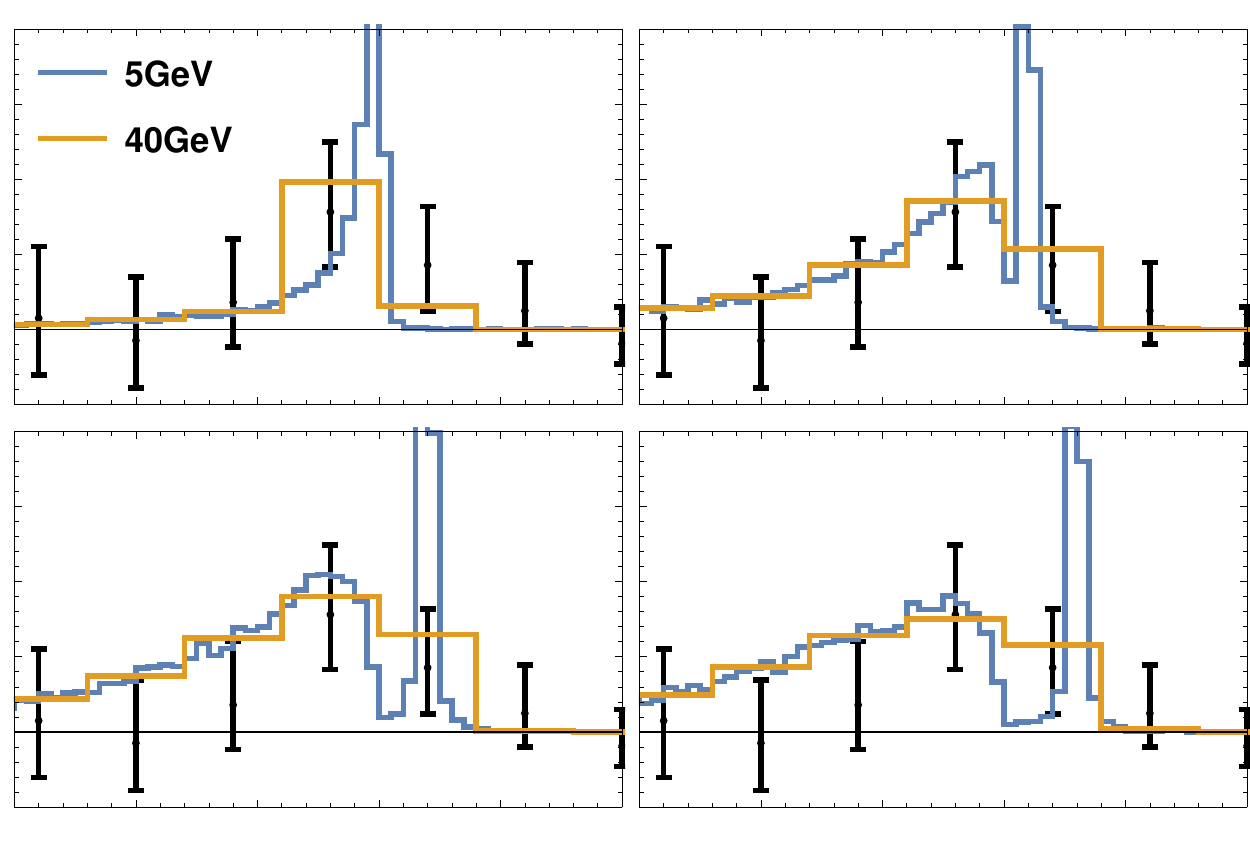}
\caption{Case B: Same as fig. ~\ref{fig:offshellA} but with $m_S=750\GeV$, $m_X=0$ and $m_P=m_S+\delta m$, for $\delta m=0,10,20,30\GeV$.}
\label{fig:offshellB}
\end{figure}

\begin{figure}
\includegraphics[width=0.5\textwidth]{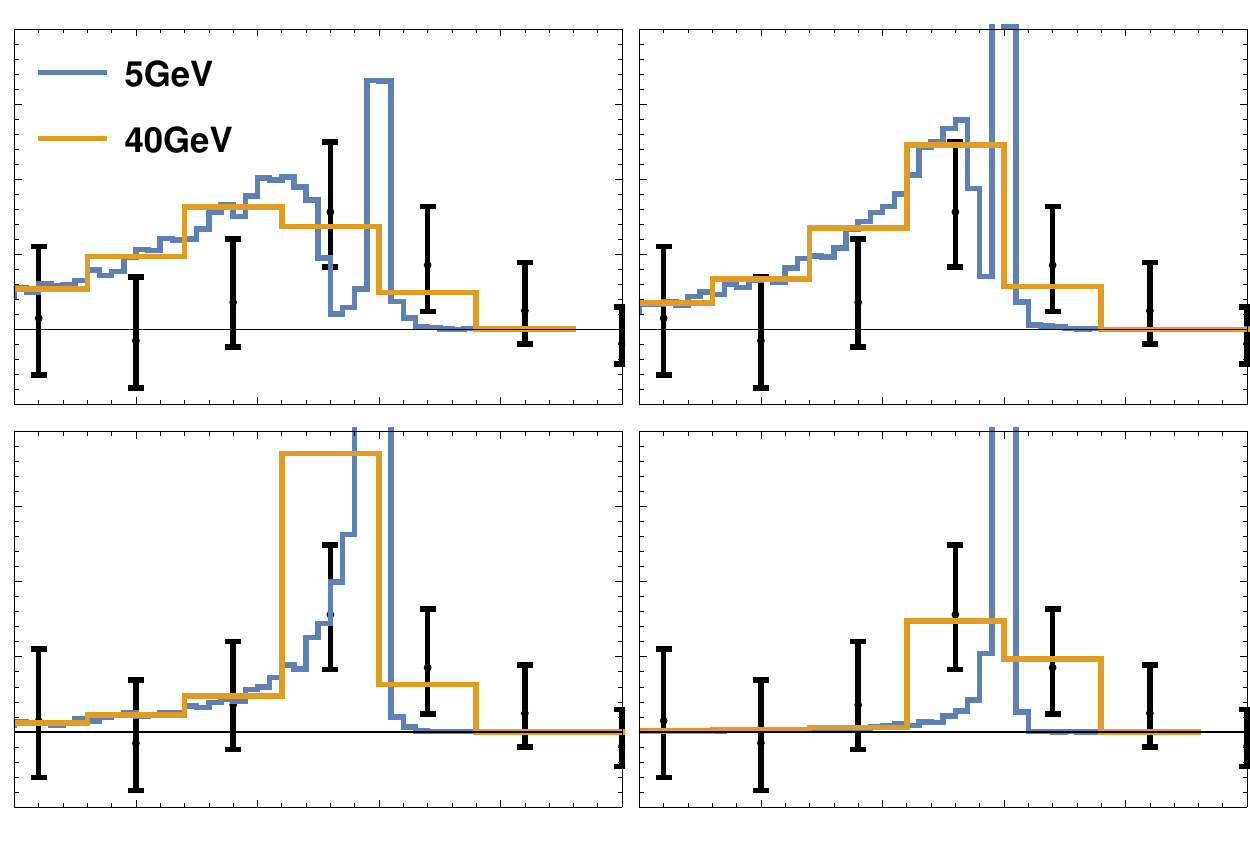}
\caption{Case C: Same as fig.~\ref{fig:offshellA} but with $m_P=750\GeV$, $m_X=0$ and $m_S=760-\delta m$, for $\delta m=0,10,20,30\GeV$.}
\label{fig:offshellC}
\end{figure}

The ATLAS results marginally prefer a resonance with a non-negligible
width (locally 3.6 vs 3.9 $\sigma$). While composite dynamics can generate a sizeable diphoton partial width $\Gamma_{P\to \gamma\gamma}$ it is hard to account for a $\Gamma_P \sim 0.05 m_P$. If such a total  width is corroborated the $P$ could be produced off-shell and mimic a finite width for $m_V<m_P+m_X$ \cite{An:2015cgp}.
In \fig{fig:offshellA}, case A, we show the diphoton distribution of a slightly off-shell $P^*$ produced together with a light (pseudoscalar) state $X$ for $m_P=m_S=750\GeV$  and $m_X=0,10,20,30\GeV$. 
In \fig{fig:offshellB}, case B, we take $m_X= 0$  and force $P^*$ off-shell from requiring $m_P=m_S+\delta m$ and $m_S=750$ GeV with  $\delta m=0,10,20,30\GeV$ while in \fig{fig:offshellC}, case C, we fix $m_P=750\GeV$  and take $m_S=760-\delta m$, again for $\delta m=0,10,20,30\GeV$. 
In the three cases it can be seen that a broad distribution which tracks the ATLAS data reasonably with the 40 GeV experimental binning is achieved. 
The second peak, seen with the finer 5 GeV binning, emerges when the parent $S^*$ goes off shell while $P$ is on shell. 
It cannot be resolved with current data, but with sufficient future LHC data this can serve as a diagnostic of the off-shell processes considered here. 

To compute the distributions, the simplified Lagrangians in \eq{eq:lag} were implemented in the UFO format \cite{Degrande:2011ua} using the Feynrules package~\cite{Alloul:2013bka}.
The Yukawa coupling is set to $y_t=0.1$ and an effective gluon operator $S G_{\mu\nu}^a G^{\mu\nu,a}/\Lambda_G$ with $\Lambda_G=6.82\times 10^4$ was implemented to reproduce the corresponding production cross section induced by a top loop, although the full loop structure in general would lead to modification in the distribution.
We used a fiducial scale $\Lambda_S=750 (1000)\GeV$ for the distributions in cases A and B (C). The correct values to reproduce the required branching ratios to diphotons are larger and shown in the dotted lines in \fig{fig:mvXLV} for cases B (orange dotted line, on the horizonthal axis $M$ refers to $m_P$) and C. The cross sections to diphotons have been rescaled accordingly. The correction to the total widths used in the figures are negligible. 
We performed a simple parton level analysis implementing the kinematical cuts used in the ATLAS analysis: $E_T^{\gamma_1}>40\GeV$, $E_T^{\gamma_2}>30\GeV$, $E_T^{\gamma_1}/m_{\gamma\gamma}>0.4$ and $E_T^{\gamma_2}/m_{\gamma\gamma}>0.3$, where $\gamma_{1,2}$ are the leading (subleading) photons in transverse energy, $E_T$.
Below we discuss explicit underlying gauge theories which in the large-$N$ limit realize this \cite{Sannino:2003xe}.

To summarize our primary parameter target in scenario 2 with a scalar parent $S$ are 
\begin{itemize}
\item $5 \times10^{-3}\lesssim  (y_{Pt})^2 \lesssim 0.7$, 
\item   $1 \gtrsim  {\rm Br}[S \to P X\to \gamma\gamma X]\gtrsim 5 \times10^{-3}$,
\item $\delta m_{SP} \equiv m_S-m_P \ll m_P$ . 
\end{itemize}

\begin{figure}
\includegraphics[width=0.5\textwidth]{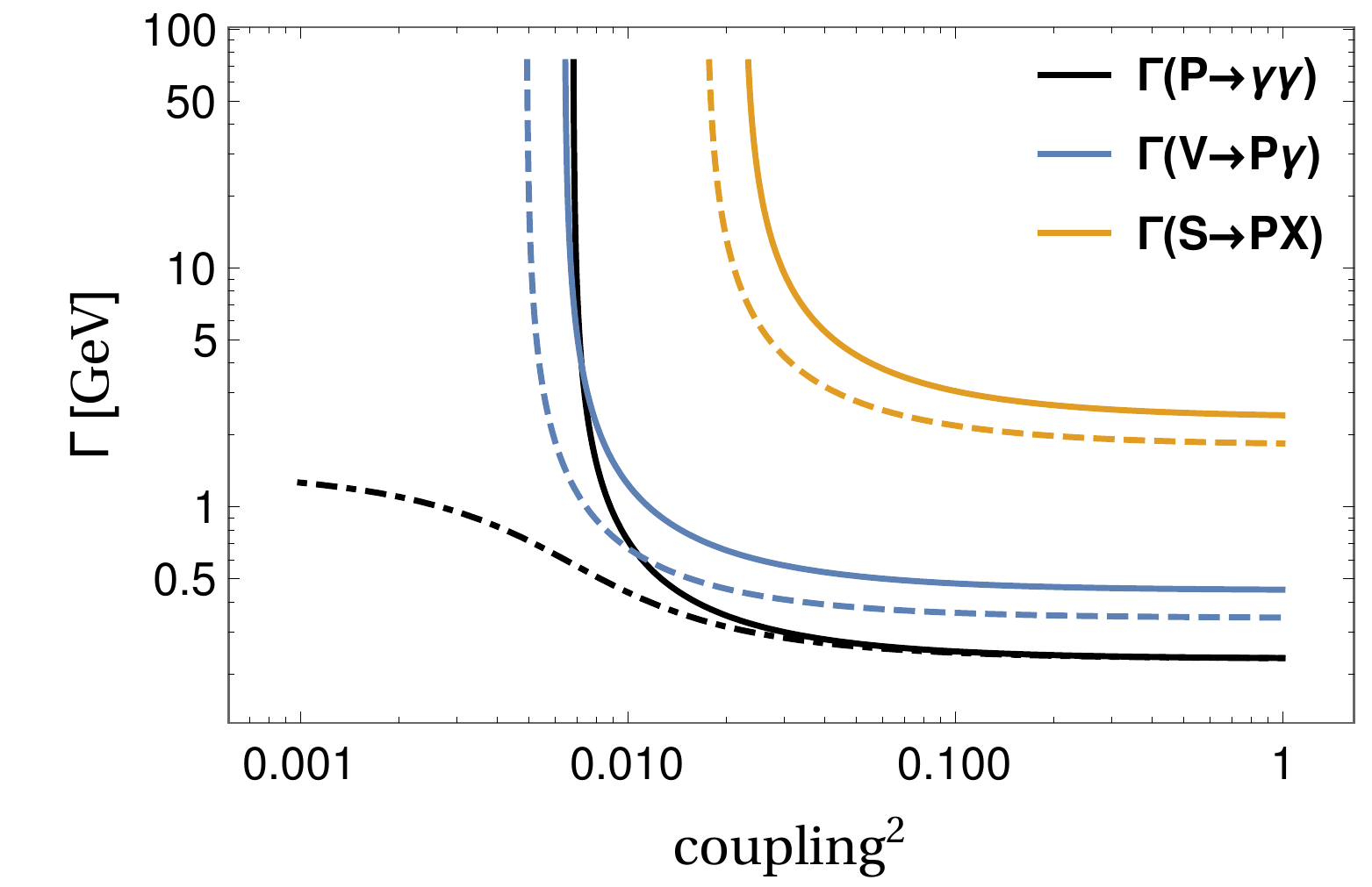}
\caption{Effect of photon fusion on the required partial widths. The dashed lines includes the photon contribution with $\Lambda_P=10\TeV$ while the solid line neglect it. For $P\to\gamma\gamma$ we show the effect of neglecting photon fusion in the solid line, while the dot-dashed one is the usual prediction with photon fusion included. We used $m_R=825\GeV$ and $r=1.64$.}
\label{fig:photonfusion}
\end{figure}

In both $V$ and $S$ scenarios, a contribution from direct production of $P$ via photon fusion will be present, relaxing the requirements on $\Lambda_S$ and $\Lambda_V$. Contrary to scenario 1, the decay rate $P\to\gamma\gamma$ can vary almost freely, although in the models we will discuss it is more natural to have $\Gamma(P\to \gamma\gamma)\lesssim 0.3$, and $\Lambda_P\gtrsim 10\TeV$. In \fig{fig:photonfusion} this effect is taken into account in the determination of the partial widths, $\Gamma(V\to P\gamma)$ and $\Gamma(S\to PX)$ and the effect is found to be small but can interplay with the parent decay topology to produce the correct excess.
For scenario 1 on the other hand, there is no freedom in choosing $\Lambda_P$ and therefore the photon fusion contribution is also fixed - we nevertheless also show the effect of neglecting it in the figure.
 
  \bigskip
  {\it \bf $P\to \gamma\gamma$ and $R\to PX$ decay widths}
 
 If $P$ arises from composite dynamics we may estimate its coupling and diphoton decay width, from loops of new strongly interacting fermions $Q$, by analogy with pseudo-scalars such as the $\pi^0, \eta$ and $\eta'$ in QCD. 

The two-photon decay width of the $\pi^0$ in QCD is given roughly by $\Gamma_{\pi \to \gamma\gamma} /m_{\pi }\simeq 6 \times 10^{-8}$ \cite{Larin:2010kq} in excellent agreement with the formula  
$\Gamma_{\pi \to \gamma\gamma} \simeq \frac{\alpha^2 m_{\pi}^2}{64\pi^3 f_\pi^2}$ with $f_\pi=93$ MeV. This formula includes the factor ${\rm Tr}[\tau_3 Q^2]=1$ where $Q$ is the charge matrix of the $u,d$ fermions and the trace includes color. For the iso-singlet $P$, similar to the $\eta$ and $\eta'$ in QCD, this becomes a sum over charges squared rather than a difference.  
For a discussion of the $\eta'$ in QCD we refer to \cite{Shore:1991np}.

Up to non-perturbative factors we therefore estimate the diphoton decay width of an isosinglet $P$ via  
\bea
\frac{\Gamma_{P\to \gamma\gamma}}{m_P} &\sim& 
 \frac{\alpha^2 m_{\pi}^2}{64\pi^3 f_\pi^2} \frac {m_P^2}{f_P^2} ( \sum_Q d(R_Q) e_Q^2)^2 
 \\
 &\sim & 3 \times10^{-8}  \frac {m_P^2}{f_P^2} ( \sum_Q d(R_Q) e_Q^2)^2 
 \eea
where we sum over the the strongly interacting constituents $Q$ of $P$, having factored out the dimension of the representation $d(R_Q)$ under the strongly interacting gauge group.

\bigskip
Thus we can get a large  diphoton decay rate to mass ratio  $\Gamma_{P\to \gamma\gamma}/m_P$ with respect to the $\Gamma_{\pi\to\gamma\gamma}/m_\pi $ if the ratio of mass to decay constant $m_P/f_P$ is large and if there is a sufficiently large number of underlying fermionic degrees of freedom. Below we give explicit examples. 
From the above estimates and \eq{Eq:partials} we identify the scale  $\Lambda_P\sim \frac{6\times 10^3 \, f_P}{ \sum_Q d(R_Q) e_Q^2}$.

Similarly we may estimate the decay rate of $V$ into $P\gamma$, from the analogous decay rates  $\omega\to \pi^0\gamma/\eta\gamma$  in QCD. From $\Gamma(\omega\to\pi^0\gamma)/m_{\omega}\sim 10^{-3}$ we find 
\bea
\frac{\Gamma_{V\to P\gamma}}{m_V} &\sim& 8 \frac{\Gamma(\omega\to\pi^0\gamma)}{m_{\omega}} \frac{f_\pi^2}{m_\omega^2}  \frac{m_V^2}{f_P^2} (\frac{\delta m_{VP}}{m_V})^3 \nonumber
\\
& \sim& 10^{-4}  \frac {m_V^2}{f_P^2} (\frac{\delta m_{VP}}{m_V})^3 \ .
\label{eq:VtoPa}
\eea
This may be compared to the minimal unavoidable decay width of $V$ into $b$-quarks, $\Gamma_{V\to bb}/m_P\gtrsim 10^{-4}$ from the required production couplings in this scenario. 
Correspondly $\Lambda_V\sim 32\,f_P$ in Eq.~(\ref{Eq:partialsV}).  Because of the strong suppression of the branching from the phase space factor $(\frac{\delta m_{VP}}{m_V})^3$ it is hard to get sufficient diphoton cross-section from this QCD-scaled estimate without additional hard activity in the final state: e.g. $\frac{\Gamma_{V\to P\gamma}}{m_V}\sim 10^{-4}$ for $\delta m_{VP}=2 f_P =150$ GeV. 

We finally consider the case of a spin-0 parent resonance $S$.  From eq.~(\ref{Eq:partials}) we have 
\bea
\frac{\Gamma_{S\to P X}}{m_S} &\simeq&   \frac{\Lambda_S^2 \delta m_{SP} }{8\pi m_S^3} 
\\
 &\simeq & \frac{\sin^2(\theta) }{16\pi} \frac{ m_S \delta m_{SP}}{f_P^2   }
\eea
In the last line we have taken $\Lambda_S \sim \frac{m_S^2}{f_P}\sin(\theta)$, by assuming a vertex of the form $\mathcal{L}= \frac{m_S^2}{f_P}  SPP$ (equivalent to the SM linear sigma model) and by introducing an angle $\theta$ that parameterizes the mixing between $P$ and $X$. 
This can be compared to the minimum width $\Gamma_{S\to tt}/m_S\gtrsim 5 \times 10^{-4}$ arising if $S$ is produced at a level of $\sigma_S\sim 5$ fb purely via a top-induced gluon fusion. Notably the smaller phase space suppression, as compared to $V\to P\gamma$ allows to achieve the cross-section more easily.

 \bigskip 
 {\it \bf Model Frameworks:}
 
 We focus on minimal composite models of EWSB as realizations of our above scenarios. Minimal we here take to mean the sector breaking EW contains a few EW doublets of strongly interacting fermions without QCD colour. Additional interactions are required to generate the SM fermion masses and Yukawa couplings of the composite states to the SM fermions. This can be done via 'ETC' interactions where 4-fermion operators bilinear in the SM fields  \cite{Eichten:1979ah,Dimopoulos:1979es} generate the required couplings, 'partial fermion compositeness' where the 4 fermion operators are linear in the SM fields \cite{Kaplan:1991dc} or by coupling the strong sector to fundamental scalar fields with Yukawa couplings  \cite{'tHooft:1979bh}.
 
A low compositeness scale $f_P< v_{EW}=175$ GeV where $f_P$ is the analogue of $f_\pi$ in QCD can lead to a large photon decay rate of $P$. This arises in composite models with multiple fermion representations \cite{Lane:1989ej} or when the composite sector is 
 coupled to, and induces vacuum expectation values for, fundamental scalars \cite{'tHooft:1979bh}. The 125 GeV scalar may then be (partially) composite and identified with one of these scalars.  

 As  explicit examples of scenario 1 and the signal process
 \bea
\sigma (pp \to P\to 2\gamma) 
  \eea
 we first consider the minimal  $SU(3)_S$ MWT model (or Next to Minimal Walking Tecthnicolor) model \cite{Sannino:2004qp} which features a single doublet of Dirac fermions $Q=(U,D)$ in the 2-index symmetric representation of $SU(3)$ with $d(R_Q)=6$ and charges $e_U=-e_D=\frac{1}{2}$ leading to $\sum_Q (d(R_Q) e_Q^2)^2=9$. Assuming $f_P/v_{\rm EW} \sim 0.5$  we find a diphoton decay width corresponding to $\frac{\Gamma_{\gamma\gamma}(P)}{m_P} \sim 3 \times 10^{-5}$. 
 
Another minimal model is a variant of the $SU(2)_{\rm Adj}$ MWT model, or simply MWT model \cite{Sannino:2004qp}, featuring a single doublet of Dirac fermions but in the adjoint representation of $SU(2)$ with $d(R_Q)=3$. Due to a Witten global anomaly this model requires an additional heavy lepton doublet $(\xi, \nu_\xi)$ in the spectrum and a gauge anomaly free charge assignment is \cite{Dietrich:2005jn}
 \bea 
 Q(U)=1, Q(D)=0, Q(\xi)=-2, Q(\nu_\xi)=-1, 
 \eea
If we assume late Yukawa couplings of the leptons to $P$ this leads to $\sum_Q (d(R_Q) e_Q^2)^2=64$. Assuming $f_P/v_{\rm EW}=0.4$ we find $\Gamma_{\gamma\gamma}(P)/m_P\sim 2\times 10^{-4}$ and with a top coupling $ y_{Pt}^2 \lesssim 0.1$ a signal cross-section of $\sigma_{\gamma\gamma}\gtrsim 5$ fb can be accommodated see \fig{fig:S1ytXGamma}.   
 
 Interestingly, by naive scaling of QCD, and identifying $P$ with the analogue of the $\eta'$ state such a decay constant $f_P$ would correspond to a $P$ mass of about 750 GeV.

While it is clear that it is not {\it immediate} to explain the observed signal cross-section without introducing new colored states to enhance production of $P$ we have here demonstrated that it is possible, even in relatively minimal models of dynamical EWSB. In models with more strongly interacting fermions carrying color, it is certainly possible \cite{Matsuzaki:2015che} to get a sufficient rate. Alternatively the composite $P$ can be coupled to new vector like coloured fermions \cite{Molinaro:2015cwg} or to heavy quarks responsible for (Witten) anomaly cancellation \cite{Antipin:2010it}.
 
 \bigskip

 As  explicit examples of scenario 2 we first consider the case of a spin-1 parent and the signal process 
 \bea
 \sigma (pp \to V \to P\,  \gamma_{\rm soft} \to \gamma\gamma \, \gamma_{\rm soft})  . 
 \eea
 
A relatively small mass splitting $\delta m_{VP}$ of the vector/pseudo-scalar system $V-P$ may be accidental. In composite dynamics it arises if $V$ and $P$ share one or more constituent fermions which are heavy with respect to the confinement scale generating the bound state.  A well known example of this is the vector-scalar mass degeneracy in the heavy quark limit of QCD.  
 
In general the symmetries of the SM allow an interaction term 
\bea 
\mathcal{L}=- \frac{\epsilon}{4}
\widehat{V}^{\mu\nu} \widehat{B}_{\mu\nu}
\label{Eq:mixing}
\eea
where  $\widehat{V}^{\mu\nu} $ is the field strength tensor of the interaction eigenstate spin-1 resonance $\widehat{V}$ and $\widehat{B}_{\mu\nu}$ is the interaction eigenstate hypercharge field. This mixing term will couple the mass eigenstate $V^{\mu}$ roughly universally to the fermion currents with a coupling of the size $\epsilon g' Y(f)$ where $g'$ is the hypercharge coupling and $Y(f)$ is the hyper charge of the fermion $f$. 

Such a kinetic mixing may equivalently be rewritten as a mass mixing of interaction eigenstates. 
However, if $V$ is composite this mass mixing with the SM weak bosons does not arise for certain anomaly free choices of weak charges of the underlying fermions \cite{Chivukula:1989rn}.  An explicit example is the $SU(3)_S$ MWT model above as shown explicitly in the appendix of \cite{Foadi:2007ue}.
Instead $V$ may still have 'direct' couplings to the SM fermion currents $J_{f , \mu}$ induced via higher dimensional operators.  If $V$ is a composite bilinear of the form $V^\mu \sim \bar{Q}\gamma^\mu Q \equiv J_Q^\mu$, where $Q$ represents the constituents of $V$, such a coupling arises via the dimension 6 operator: 
\bea
\frac{1}{\Lambda_b^2} J_Q^\mu J_{b , \mu} \sim \frac{1}{\Lambda^2_b} \bar{Q}\gamma^\mu Q  \, \bar{b}\gamma_\mu b 
\eea
where $\Lambda_b$ is related to the mass scale of the interactions mediating the 4-fermion operators.

The effective Lagrangian for the composite states can be described via a chiral (vector resonance) Lagrangian with the goldstone bosons contained in a non-linear sigma field $U$ such that we can identify 
\bea 
J_Q^\mu &\to & f_P^2 \rm{Tr}(i D^\mu U  \, U^\dagger )   \to f_P^2 \tilde{g} V^\mu 
 \\
  D^\mu U &=& \partial^\mu U - i \tilde{g} V^\mu U
\eea
where the nonlinear sigma field on the vacuum is normalized to $\langle U \rangle =1$ and $\tilde{g}\lesssim 4\pi$ is the effective strong interaction coupling. In QCD $\tilde{g}$ would be related to $g_{\rho \pi\pi}$ where we identified the 'pion' decay constant with $f_P$.

Then from the above composite operator we find 
\bea 
\frac{1}{\Lambda^2_b} J_Q^\mu J_{b , \mu}  \to  \frac{f_P^2}{\Lambda^2_b}  \tilde{g} V^\mu  J_{b , \mu} \, ,  
\eea
and require 
$ \frac{\tilde{g} f_P^2}{\Lambda^2_b}\gtrsim 4.2 (5.6) \times 10^{-2} $ for $\delta m_{VP}=75 (175)\GeV$ to get a production cross-section of at least 5 fb. 
Even for larger couplings, \emph{e.g.} $\frac{\tilde{g} f_P^2}{\Lambda^2_b}\gtrsim 0.1$ the scale varies little around
 $f_P\sim \Lambda_S/32 \sim 10 (20)\GeV$ for mass splitting $\delta m_{PV}=75 (175) \GeV$, which is very small - even if the constituent quarks have large masses. The corresponding values of $\Lambda_b\lesssim \sqrt{\frac{\tilde{g}}{0.1}} f_P\sim \sqrt{\tilde{g}}\, 32(63)\GeV$ are around the weak scale $v_{\rm EW}\simeq 175$ GeV for $\tilde{g}\sim 4\pi$.

These interactions may e.g be induced via ETC type interactions of the form 
\bea
\frac{1}{\Lambda^2_b} \bar{Q}_L \sigma^\mu b_L Q_R\sigma^\mu \bar{b}_R  \, , 
\eea
where we identify $\Lambda_b \sim M_{ETC, b}/g_{ETC}$ with the ratio of mass and coupling of an exchanged ETC gauge boson. 
Upon Fierz rearrangement and condensation of $Q$'s this leads to a mass term for the $b$-quark and will include interaction terms of the form 
\bea
\frac{c_1}{ \Lambda^2_b}\bar{Q}\gamma^\mu Q  \, 
\bar{b}\gamma_\mu b + \frac{c_2}{ \Lambda^2_b}\bar{Q}\gamma^5 Q  
\, \bar{b}\gamma_5 b + ...
\eea
where $c_{1,2}$ are $O(1)$ numbers. 
Getting a sufficiently large coupling requires $M_{ETC, b}\sim g_{\rm ETC}\sqrt{\tilde{g}}\, 32(63)\GeV$ which still for $\tilde{g}\sim 4\pi, g_{\rm ETC}\sim 4$  is low for typical ETC models.

Again the scenario is borderline able to account for the observed excess of photons at LHC. In particular the analysis has neglected Clebsch-Gordon coefficients from the fierzing of group structure matrices in the ETC interactions. We leave it as a proof of principle that an underlying composite model can be constructed to realize our scenario 2.

%

 We finally consider the case of a spin-0 parent resonance $S$ and the signal process 
 \bea
 \sigma (pp \to S \to P X \to \gamma\gamma X)  . 
 \eea
We identify $S$ with a (partially composite) scalar with sizable Yukawa couplings either from mass mixing with a fundamental scalar as in  \cite{'tHooft:1979bh} or via 4-fermion interactions \cite{Eichten:1979ah,Dimopoulos:1979es,Kaplan:1991dc}. 
Like the $\sigma$ resonance in QCD we expect a significant coupling to pseudo scalar pairs $\mathcal{L}=\Lambda_S  SPP$ with $\Lambda_S \sim \frac{m_S^2}{f_P}$ leading to an $O(1)$ width for $\Gamma_{S\to PP}$ for massless $P$s. 
With $m_P\simeq 750$ GeV the decay mode is closed kinematically but we imagine that $P$ mixes with lighter state(s) of the same quantum numbers. An interesting possibility is the scenario in \cite{Frandsen:2011kt} where a new few GeV scale of a neutral composite sector is present and provides DM candidates. If $X$ denotes a peudoscalar state then $P$ can mix with it.

At the minimal Yukawa coupling $y_{St}$ required to reproduce the di-photon excess exclusively via the top-loop induced process, the scale $\Lambda_S$  as well as the required partial width of $S\to PX$ diverges, and $f_P$ tends to vanish. Near this low values of $y_{St}$ photon fusion plays an important role in the production of the $P$ state, as discussed in scenario 1. Fixing the previously introduced mixing angle $\sin(\theta)=0.1$, in order to get $f_P\gtrsim 15\GeV$ it is necessary $y_{St}^2\sim 0.012$ for $\delta m_{SP}=25\GeV$ and photon fusion may still plays an important role for typical values of $\Lambda_P$.

For larger values of $y_{St}^2\gtrsim 0.1$ the top-quark dominates both the production via loop and decay widths of $S$, making $\Lambda_S$, $f_P$ and the partial width $S\to PX$ varying little in $y_{St}$. For $y_{St}=0.1$ and $\sin(\theta)=0.1$ we have $f_P=73,\,107,\,119\GeV$ for $\delta m_{SP}=25,\,75,\,175\GeV$.
We conclude that our scenario 2 is more easily achieved with a scalar parent resonance than with a vector.

Finally we note that in this scenario 2, where $P$ may have negligible couplings to SM, $P$ may still be produced directly via photon-fusion, given the sizable diphoton couplings that can be achieved.

\bigskip
{\bf  Possible signatures: }
 
In the limit where $P$ is produced off-shell a distinct feature which can identify this scenario 2 over the photon fusion, or the gluon fusion production of $P$ in scenario 1, is the double peak, or peak plus tail, structure in the diphoton signal, which can be resolved with more statistics and finer binning, see figs.\ref{fig:offshellA}-\ref{fig:offshellC}.

 Once $P$ is produced on-shell but with the additional final state(s) $X$ very soft,  it will be complicated to distinguish from additional activity in the simple gluon fusion production in scenario 1.  
Finally in the limit where the mass splitting becomes sizable the tell-tale sign of the vector realization $V\to P \, \gamma/Z$ of scenario 2 will be the the production of an additional photon or an on shell $Z$ boson reconstructing the $V$ invariant mass above the $P$ mass.

 \bigskip
  {\bf Variations:}

 In other realizations than the one considered here it is possible to motivate larger couplings of the vector resonance to $b$-quarks which in turn would allow a broader resonance $P$ while still achieving the observed signal cross-section. At the prize of increased tension with LHC8 data it is also simply possible that $R$ is produced via the light SM fermions. This would be the case in composite models of a vector $R=V$ and underlying fermion charges allowing  a direct mass-mixing between $V$ and the SM hypercharge $B$ field. 

Another interesting possibility is the regime where $P$ is extremely light such that the decay of $P$ via $V\to (P\to \gamma\gamma) \gamma$ would lead to a highly collimated diphoton pair \cite{Chala:2015cev,*Knapen:2015dap}. 
This spectrum would be possible if $P$ were a nearly exact Goldstone boson.

 \bigskip 
 {\it \bf Summary:}
 
 We have considered 2 scenarios for  explaining the recently observed excesses in diphoton final states near $750$ GeV invariant masses by the ATLAS and CMS collaborations \cite{ATLAS,CMS:2015dxe}. Our motivation was to connect the signal to models of (dynamical) electroweak symmetry breaking and flavor and to avoid having to invoke new vector-like colored states. 
 
In both scenarios we assume the excess is due to a pseudo-scalar resonance produced either directly in top-induced gluon fusion or via a parent resonance $R$ close in mass. We take the parent resonance $R$ to be either spin-1 or spin-0. With a small mass splitting between $R$ and $P$ the additional activity in the final state beyond the diphoton pair can be soft. We further assumed the spin-1 parent couples dominantly to $b$-quarks in order to have a large enhancement of the production cross-section from the 8 TeV to the 13 TeV run. While such mass degeneracies and couplings are in general ad hoc we motivated them in symmetry limits of scenarios of weak scale composite dynamics.

In all scenarios a significant coupling of $P$ to photons is induced by new (dynamically) heavy strongly interacting fermions.

If the signal is due to a pseudo-scalar produced in gluon fusion via an order 1 top Yukawa, producing a width $\Gamma_P \sim 0.05 m_P$ the pseudo-scalar must have a very significant partial width into photons. We found that this is possible even in relatively minimal, but low scale composite models of EWSB.  Thus an additional sector playing a role in EW symmetry breaking must in this case be present. 

If the signal is instead due to production of $P$ via a parent resonance  the branching of $P$ into diphotons can be $O(1)$. We discussed the possibility of a spin-1parent resonance coupled dominantly to $b$-quarks, motivated by the $\omega$ meson in QCD and by the dominant third generation couplings induced from 'Extended Technicolor'-type interactions. 
However in this case it is hard to reach the require cross-section.

Finally if the parent resonance is a scalar $S$, top couplings of $S$ might be induced due to mass mixing with the $125$ GeV Higgs, while fermion couplings of $P$ could be negligible. 
In that case we could have a very narrow resonance with a branching ratio into photons of order 1. If a large width of the excess observed in ATLAS data persists this could be due to a slightly off-shell production of $P$.

 The distinct experimental signature of the second scenario is the presence of an additional soft photon or off-shell $Z$ (for sufficiently small mass splittings) in the final state, while in the third scenario with a scalar parent, there would be additional soft hadronic activity.   
The most important motivation for our study is the connection of the diphoton excess to underlying models of dynamical EWSB.

\acknowledgements
We thank E. Molinaro, F. Sannino and N. Vignaroli for comments and discussions. 
The CP$^3$-Origins centre is partially funded by the Danish National Research Foundation, grant number DNRF90.
%
%

\bibliography{diphoton-2}
\end{document}